\documentclass{acm_proc_article-sp}

\usepackage[scaled=0.83]{beramono}

\usepackage{listings}

\lstdefinelanguage{diaspec}
{morekeywords=[1]{entity, attribute, extends, source, action, controller,
context, from, on, void, Integer, Boolean, String, enumeration, Picture,
structure, indexed, by, as, raises, get, when, provided, maybe,
gives, takes,
publish, interaction, mandatory, catch, in, frequency, always, do, nothing,
require, available, availability},
morekeywords=[2]{as, source, Bool, Int, resource, required, no, use, component},
  sensitive=true,
  morecomment=[l]{//},
  morecomment=[s]{/*}{*/},
emph={%
    srcName, actName, ctxName, ctrName%
    },emphstyle={\emph}%
}
\newcommand{\ie}{\emph{i.e.,}\xspace}
\newcommand{\etc}{\emph{etc.}\xspace}

\newcommand{\eg}{\emph{e.g.,}\xspace}

\newcommand{\ignore}[1]{}

\newif\ifshowcomments
\showcommentstrue

\ifshowcomments
\newcommand{\mynote}[2]{%
\fbox{\bfseries\sffamily\scriptsize#1} %
  {$\blacktriangleright$\textcolor{red}{\textsf{\emph{\footnotesize#2}}}$\blacktriangleleft$}}%
\else
\newcommand{\mynote}[2]{}
\fi

\usepackage{listings}
\usepackage[usenames,dvipsnames]{xcolor}
\usepackage{soul}

\definecolor{light-gray}{gray}{0.40}

\lstdefinelanguage{racket} {
  morekeywords=[1]{define, define-syntax, define-macro, lambda, define-stream, stream-lambda},
  morekeywords=[2]{begin, call-with-current-continuation, call/cc,
    call-with-input-file, call-with-output-file, case, cond,
    do, else, for-each, if,
    let*, let, let-syntax, letrec, letrec-syntax,
    define-context, define-controller, Integer, Boolean, get, when-required, when-provided,
    maybe_publish, require, submod, or/c, ->, \#\%module-begin,
    always_publish, with-syntax, define-struct/contract, syntax-case,
    define/contract,
    let-values, let*-values,
    module, provide,
    and, or, not, delay, force,
    \#`, \#',
    \#lang, implement, begin-for-syntax, rename-out,
    quasiquote, quote, unquote, unquote-splicing,
    map, fold, syntax, syntax-rules, eval, environment, query },
  morekeywords=[3]{import, export},
  alsoletter={',`,-,/,>,<,\#,\%},
  morecomment=[l]{;},
  moredelim=**[is][\color{light-gray}]{<<@<<}{>>@>>},
  moredelim=**[is][\itshape\color{OliveGreen}]{<<;<<}{>>;>>},
  morecomment=[s]{\#|}{|\#},
  sensitive=true,
}

\lstdefinestyle{box}{
  keywordstyle=\color{blue},
  commentstyle=\color{green},
}

\def\inlineracket{\lstinline[basicstyle=\ttfamily\normalsize,language=racket]}

\usepackage{cite}

\usepackage[activate={true,nocompatibility},%
final,%
tracking=true,%
kerning=true,%
spacing=true,%
factor=1100,%
stretch=10,%
shrink=10]{microtype}
\microtypecontext{spacing=nonfrench}

\usepackage{xspace}
\xspaceaddexceptions{]\}}

\usepackage{graphicx}
\usepackage{amssymb}
\usepackage[T1]{fontenc}
\usepackage{xstring}
\usepackage{textcomp}

\newcommand{\etal}{\emph{et al.}\xspace}
\newcommand{\cf}{\emph{c.f.}\xspace}

\usepackage{hyperref}
\hypersetup{%
  colorlinks,%
  citecolor=Blue,%
  linkcolor=black,%
  urlcolor=Blue}

\newcommand{\subsurl}[1]{{%
  \noexpandarg 
  \StrSubstitute{#1}{http://}{http:{//}}[\x]
  \expandafter\StrSubstitute\expandafter{\x}{https://}{https:{//}}[\x]%
  \expandafter\StrSubstitute\expandafter{\x}{-}{-\allowbreak }[\x]%
  \expandafter\StrSubstitute\expandafter{\x}{/}{\allowbreak/}[\x]%
  \x}}

\renewcommand{\url}[1]{\href{#1}{\texttt{\subsurl{#1}}}}

\makeatletter
\let\@copyrightspace\relax
\makeatother

\numberofauthors{1} 
\author{%
\alignauthor
Paul van der Walt\\
       \affaddr{INRIA Bordeaux, France}\\
       \email{\texttt{\large paul.vanderwalt@inria.fr}}%
}

\date{\today}
\title{Constraining application behaviour\\ by generating languages}

\begin{document}
\maketitle

\begin{abstract}
Writing a platform for reactive applications which enforces operational
constraints is difficult, and has been approached in various ways. In
this experience report, we detail an approach using an embedded DSL
which can be used to specify the structure and permissions of a
program in a given application domain. 
Once the developer has specified which components an
application will consist of, and which permissions each one needs, the
specification itself evaluates to a new, tailored, language.
The final implementation of the application is then written in this
specialised environment where precisely the API calls associated with
the permissions which have been granted, are made available.

Our prototype platform targets the domain of mobile computing, and is
implemented using Racket. It demonstrates resource access control (\eg
camera, address book, \etc) and tries to prevent leaking of private
data. Racket is shown to be an extremely effective platform for
designing new programming languages and their run-time libraries.  We
demonstrate that this approach allows reuse of an inter-component
communication layer, is convenient for the application developer
because it provides high-level building blocks to structure the
application, and provides increased control to the platform owner,
preventing certain classes of errors by the developer.
\end{abstract}


\section{Introduction}
\label{sec:introduction}

Among programming frameworks intended to be used by third party
developers, we see a trend towards including mechanisms restricting
access to certain features, or otherwise constraining behaviour of the
application~\cite{DBLP:books/daglib/0029092,iphonesdk}. In the case of
platforms like Android~\cite{android}, the aim is usually to protect
the user's sensitive data (\eg contact list, physical location) from
undesired use, while still giving applications access to the
resources, whether hardware or data, needed to function correctly. For
example, an email application legitimately requires access to the
Internet, but for a calculator application this should raise
suspicion.  In the case of Android, these restrictions are enforced
via run-time checks against a permissions file called the Manifest,
which the user accepts at install-time. Other frameworks are also
adopting such declarations, in various
forms~\cite{chrome-extensions,Feiler:2008:EFA:1593806}.

\subsection{Declaration-driven frameworks}
\label{sec:decl-driv-fram}

Generally speaking, we identify a class of programming frameworks in
widespread use, which we call \emph{declaration-driven
frameworks}. These frameworks are different to traditional static
programming frameworks in that they have some form of declarations as
input.  Examples abound, including the Android SDK
with its Manifest file, or the Facebook plugin SDK, both of which
require permissions to be granted \emph{a priori}. The declarations
vary greatly in expressiveness, on a spectrum from simple resource
permissions, \eg access to list of friends and the camera, to very
expressive, \eg rules for the control flow of the application, a list
of components to be implemented, \etc An example of the latter is
DiaSuite~\cite{DBLP:journals/tse/CassouBCB12}, where the individual
components of the application, as well as their subscription relations, are
laid out in the declarations. Access to resources is also granted on a
component level. These rich declarations encourage
separation of components, and provide relative clarity for the user
regarding potential information flow, when compared to a simpler list of permissions.
Diagrams with potential information flow can be extracted from the specifications, 
and presented in graphical format, for example.

\subsection{Problem}
\label{sec:motivation}

We identify a number of shortcomings with the systems mentioned
above. Most frequently, the declarations are no more than a list of
permissions checked at run-time, leaving the user guessing about the
actual behaviour of the
application~\cite{DBLP:conf/kbse/XiaoTFHM12}. The fact that these
checks are dynamic also leads to the application halting on Android if
a developer tries to access a forbidden resource. Existing static
approaches that exist, on the other hand, generally try to solve
different problems than resource access
control, for example just checking that 
all required components are implemented~\cite{DBLP:books/daglib/0029092}.

These shortcomings are addressed by DiaSuite's approach allowing static
checks on resource access, which is based on an external DSL. However,
the current implementation of DiaSuite generates Java boilerplate code
from the declarations, tailored to the specific application.  With
this generative approach, extending the declaration language would
involve modifying the standalone compiler, and in general, generated
code tends to be difficult to debug and inconvenient to work with.

On the other hand, the language building platform provided by Racket 
allows simple implementation of an embedded DSL, with all the features
of Racket potentially available to the application
developer. Providing expressive constructs for specification and
implementation of an application raises the level of abstraction, and
allows us to implement static guarantees of resource access equivalent to
DiaSuite.  Furthermore, we need not maintain parser and compiler
machinery in parallel with the framework infrastructure.

In this report we demonstrate the use of Racket's language extension
system~\cite{tobin2011languages} allowing us to easily derive tailored
programming environments from application declarations.  This
decreases effort for the application developer and gives more control
to platform owners.  To the best of our knowledge this is the first
implementation of an EDSL which itself gives rise to a tailored EDSL in
Racket. The code presented in this report is available from
\url{http://people.bordeaux.inria.fr/pwalt}.

\paragraph*{Outline}
After giving a brief overview of related work in 
Sec.~\ref{sec:related-work-1}, we introduce the platform we have chosen
as the basis of our prototype, as well as the example
application to be implemented  in Sec.~\ref{sec:case-study}.
In Sec.~\ref{sec:implementation} we show how a
developer using our system would write the example application.
Sec.~\ref{sec:framework} goes into detail about how the declarations
unfold into a language extension, and finally in
Sec.~\ref{sec:discussion} we discuss strengths and weaknesses of the
approach using Racket. Our conclusions are presented in
Sec.~\ref{sec:conclusion}.

\section{Related work}
\label{sec:related-work-1}

The work most closely resembling ours is
DiaSuite~\cite{Cassou:2011:LSA:1985793.1985852}, since it is the
inspiration for our approach. The relative advantages and
disadvantages are thoroughly dealt with in Sec.~\ref{sec:case-study}.

Other than that, it seems there is not much literature on the
generation of frameworks, although to varying degrees frameworks which
depend on declarations are becoming ever more widely
adopted~\cite{android,chrome-extensions}. These generally address
demonstrated threats to user
safety~\cite{elish2013static,wei2012permission,mann2012framework,stevens2012investigating,gibler2011androidleaks}.

Many approaches have been proposed to address these leaks, such as 
parallel remote execution on a remote VM where a dynamic taint analysis is 
running~\cite{enck2014taintdroid}. This naturally incurs its own privacy 
concerns, as well as dependence on a connection to the VM.
Another approach which bears similarity to our aim, is the work
by Xiao \etal~\cite{DBLP:conf/kbse/XiaoTFHM12}, which restrains
developers of mobile applications to a limited external DSL based on
TouchDevelop~\cite{touchdevelop-book}, from which they extract
information flow via static analysis. This information is then
presented to the user, to decide if the resource usage seems
reasonable. This is a powerful and promising approach, but we believe
that it is preferable to declare information flow paths \emph{a
priori} and constrain the developer, than having to do a heavy static
analysis to extract that same information -- especially since it means
a developer cannot use a general-purpose language they are already
familiar with, but must learn a DSL which is used for every aspect of
the implementation.

Compared to these alternatives, providing a ``to\-wer of
lang\-ua\-ges''-\allowbreak style solution~\cite{JFP:8573399} seems be a good
trade-off between restrictions on the developer and versatility of the
implementation.

\section{Case study}
\label{sec:case-study}

DiaSuite, the model for our prototype, is a declaration-driven
framework which is dedicated to the \emph{Sense/Compute/Control}
architectural style described by Tay\-lor \etal
\cite{taylor2009software}. This pattern ideally fits applications that
interact with an external environment.  SCC applications are typical
of domains such as building automation, robotics, avionics and
automotive applications, but this model also fits mobile computing.

\subsection{The Sense-Compute-Control model}
\label{sec:sense-comp-contr}

\begin{figure}
\centering
  \includegraphics[width=0.75\linewidth]{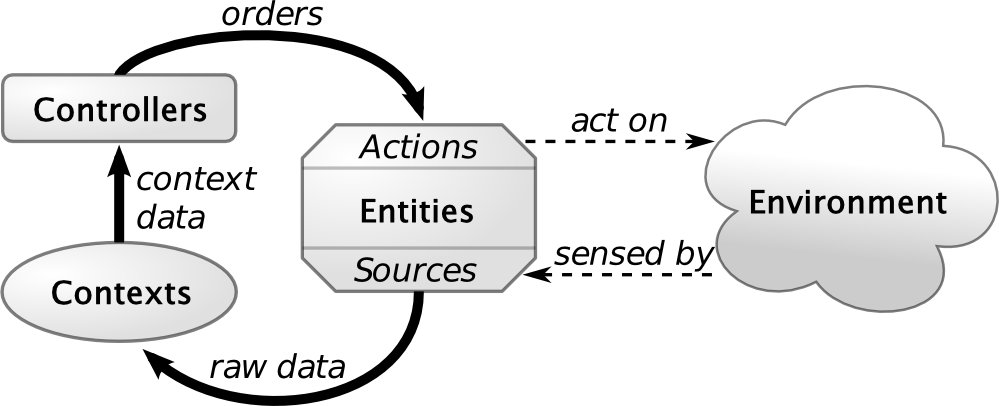}
  \caption{The \emph{Sense/Compute/Control} paradigm. Illustration
    adapted from~\cite{DBLP:journals/tse/CassouBCB12}.}
  \label{fig:sccpattern}
\end{figure}

As depicted in Fig.~\ref{fig:sccpattern}, this architectural pattern
consists of three types of components: (1)~\emph{entities} correspond
to managed\footnote{Managed resources are those which are not
  available to arbitrary parts of the application, in contrast to
  basic system calls such as querying the current date.}  resources,
whether hardware or virtual, supplying data; (2)~\textit{context
  components} process (filter, aggregate and interpret) data;
(3)~\textit{controller components} use this information to control the
environment by triggering actions on entities. Furthermore, all
components are reactive. This decomposition of applications into
processing blocks and data flow makes data reachability explicit, and
isolation more natural. It is therefore well-suited to the domain of
mobile computing, where users are entrusting their sensitive data to
applications of dubious trustworthiness.

\begin{figure}
  \centering
\begin{lstlisting}[language=diaspec]
Declaration -> Resource | Context | Controller
Type        -> Bool | Int | String | !\ldots!
Resource -> (source srcName | action actName) as Type
Context  -> context ctxName as Type CtxtInteract
CtxtInteract -> when ( required GetData?
                     | provided (srcName | ctxName)
                         GetData? PublishSpec)
GetData       -> get (srcName | ctxName)
PublishSpec   -> (always | maybe) publish
Controller    -> controller ctrName ContrInteract
ContrInteract -> when provided ctxName do actName
\end{lstlisting}
  \caption{Declaration grammar. Keywords are in bold, terminals in
italic, and rules in normal font.}
  \label{fig:diacoregrammar}
\end{figure}

\subsection{Declaration language}
\label{sec:declaration-language}

The minimal declaration language associated with DiaSuite is presented in Fig.~\ref{fig:diacoregrammar}. 
It is adapted from \cite{DBLP:journals/tse/CassouBCB12}, keeping only essential constructs. 
An application specification is a list of \texttt{Declaration}s. 
Resources (such as camera, GPS, \etc)
are defined and implemented by the platform: they are
inherent to the application domain.
Context and controller declarations include interaction
contracts~\cite{Cassou:2011:LSA:1985793.1985852}, which
prescribe how they interact.  A context can be activated by either another component
requesting its value (\texttt{when required}) or a publication of a value by another component (\ie \texttt{when provided \emph{component}}).
When activated, a context component may be allowed to pull data
(denoted by the optional \texttt{get}). Note that 
contexts which may be pulled from must have a \texttt{when required} contract. Finally, a
context might be required to publish when triggered (defined by
\texttt{Pub\-lish\-Spec}). Note that \texttt{when required} contexts
have no publish specification, since they are only activated by pulling, and hence return their values 
directly to the component which polled them.  When activated,
controller components can send orders, using the actuating
interfaces of components they have access to (\ie \texttt{do
\emph{actName}}), for example printing text to the screen or sending an email.

DiaSuite  compiles the declarations, written in an external DSL, into
a set of Java abstract classes, one for each declared component, plus an
execution environment to be used with the classes which extend them.
The abstract classes contain method headers which are derived from
the interaction contracts, and constrain the input and output of the developer's
implementation of each component. Additionally, access to resources is
passed in as arguments to these methods, so that the only way a
developer may use a resource is via the capability passing method from
the framework.

This approach allows advantages such as static checks by the Java
compiler that the application conforms to the declarations. From these
declarations it directly follows which sensitive resources components
should have access to, giving the application developer a much more
concise API to work with.  For example, if a component only has access
to the network, it need not have the API for dealing with the camera
in scope. This is in contrast to Android, where all system API calls
are always available, increasing the amount of information the
developer must keep in mind.  The disadvantage is that this is an
external DSL, and thus requires a separate compiler to be
maintained. It also implies less versatility, having to re-invent the wheel,
and a symbolic
barrier, as argued by Fowler~\cite{fowler2010domain}.

\subsection{Example application}
\label{sec:our-example-application}

\begin{figure}
  \centering
  \includegraphics[width=0.50\linewidth]{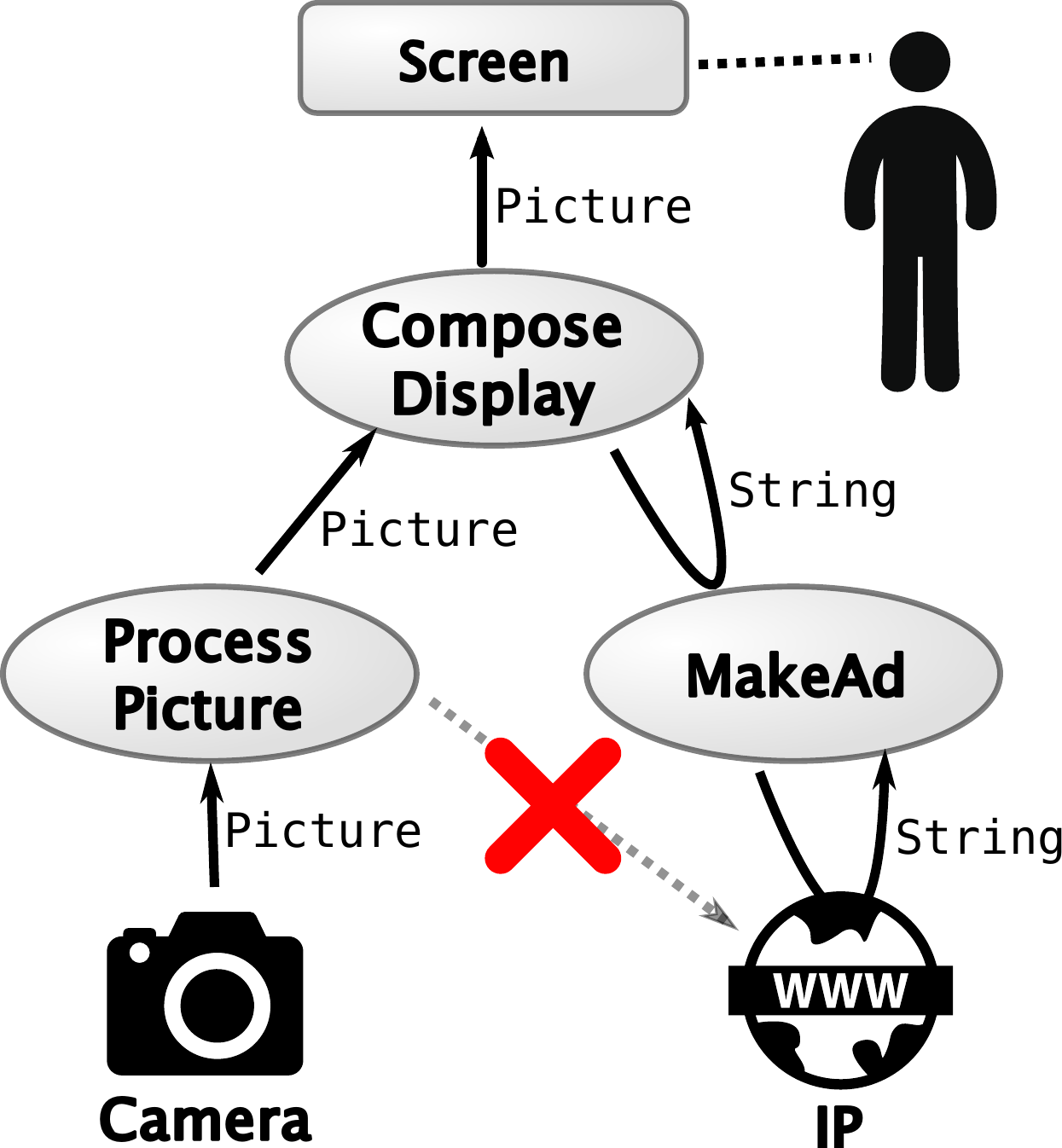}
  \caption{Simplified schematic of example application's design. We do not want the picture to be able to
leak to the WWW. Note that pull requests (the curved arrows) are not parameterised,
and are only used to return values.}
  \label{fig:schematic-application}
\end{figure}

As our running example, we use a prototype mobile application. We
pretend that it is distributed for free, supported by
advertisements. It allows the user to capture pictures and then view
them with a colourful filter (see
Fig.~\ref{fig:schematic-application}). An advertisement will be
downloaded from the Internet, but we would like to prevent the
developer from being able to leak the picture (which is private) to
the outside world, whether intentionally or by using a malicious
advertisement provider. It has been shown that this is in fact a
threat: frequently, included third party advertisement libraries try
to exfiltrate any private data to which they are able to get
access~\cite{stevens2012investigating}.

From the specification it follows that it should be impossible for the
picture to leak to the Web, since the bitmap processing component is
separate from the advertisement component. 

\section{Implementation of example}
\label{sec:implementation}

\begin{figure}
  \centering
  \includegraphics[angle=-90,width=\linewidth,trim=0 0 1cm 0,clip=true]{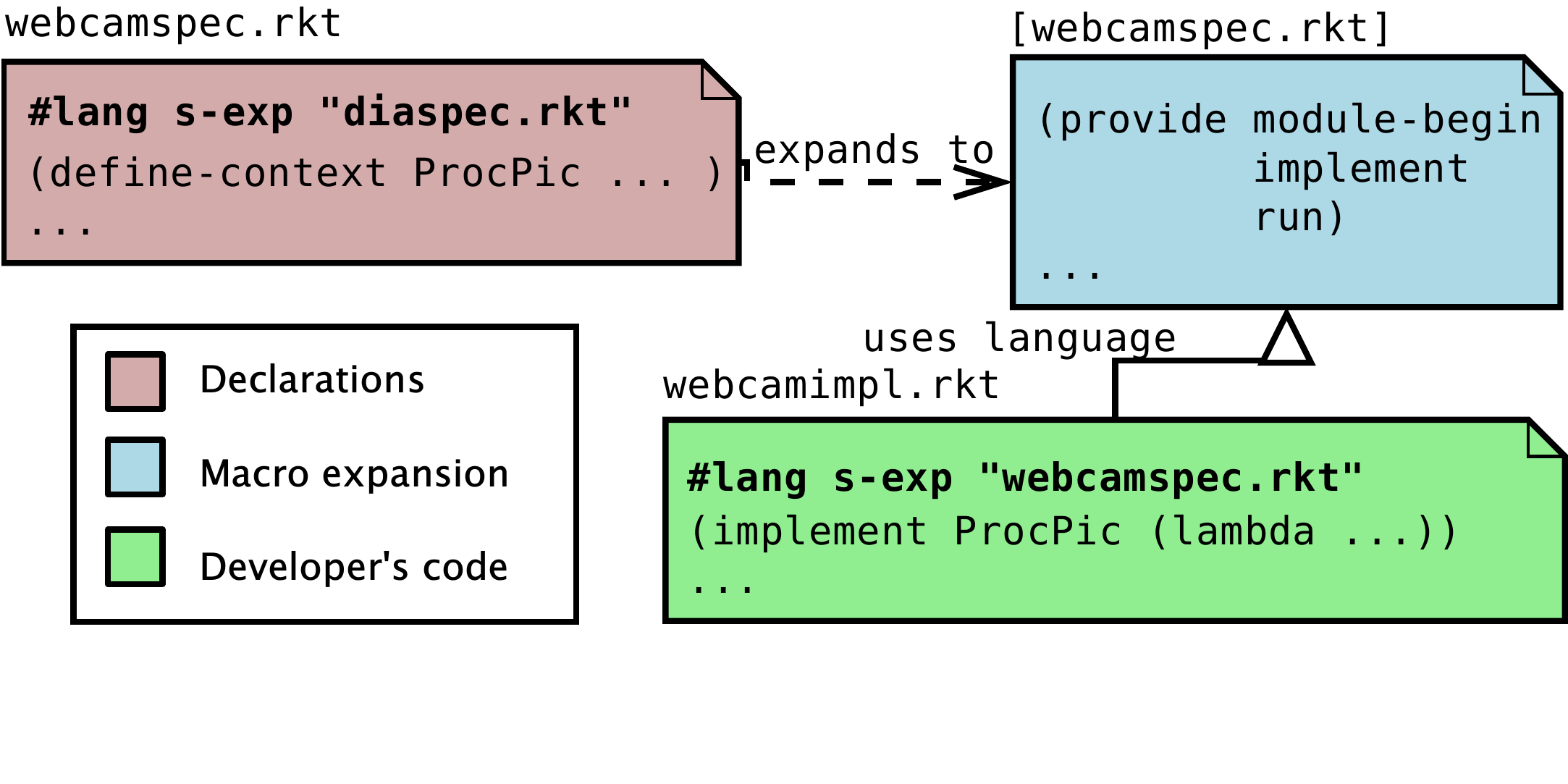}
  \caption{The prototype's architecture. Provided
declarations are transformed into a tailored language for the
implementation. The \texttt{implement} macro gets cases for each
declared component.}
  \label{fig:racket-global}
\end{figure}

Inspired by the DiaSuite approach, where a framework is generated from
the specifications, the first step in our implementation is to provide
an embedded DSL for writing specifications. It should include
constructs for defining contexts and controllers, according to the
grammar in Sec.~\ref{sec:case-study}.  As illustrated in
Fig.~\ref{fig:racket-global}, when the specifications are evaluated,
they in turn form a language extension which should be used to
implement the application. The programming environment that is thus
created provides the developer with tailored constructs for the
application that is to be built, including an API precisely matched to
what each component may do. In our prototype, we consider the advert developer 
and application developer as potentially the same, since we expect the advertisement
library to provided in the form of a snippet of code that will be 
included along with the rest of the implementation code. This way, the advertisement
code does need to be specially analysed, but is subject to the same constraints
as any other code provided by the developer. That is, it can only access
entities specified in the declarations.

\subsection{Example specifications}
\label{sec:specification-dsl}

The specifications as rendered in Racket, for our example application,
are shown in Fig.~\ref{fig:racket-spec}. The syntax closely matches 
the DiaSuite declaration language previously introduced, and 
reflects the graphical representation of the application in Fig.~\ref{fig:schematic-application}.

\begin{figure}
  \centering
\begin{lstlisting}[language=racket]
;;; Specifications file, webcamspec.rkt
#lang s-exp "diaspec.rkt"
(define-context MakeAd String [when-required get IP])
(define-context ProcessPicture Picture
  [when-provided Camera always_publish])
(define-context ComposeDisplay Picture     !\label{code:compose-display-spec}!
  [when-provided ProcessPicture get MakeAd
     maybe_publish])
(define-controller Display
  [when-provided ComposeDisplay do Screen])
\end{lstlisting}  
  \caption{Complete declarations of the example application, 
in Racket prototype.}
  \label{fig:racket-spec}
\end{figure}

\subsection{Semantics of declarations}
\label{sec:semant-decl}

In this section, we explain the semantics of each term, from the
point of view of the application developer.

The keywords \texttt{define-\allowbreak context} and
\texttt{define-\allowbreak controller} are available for specifying
the application, and upon evaluation, will result in a macro
\texttt{implement}, for binding the implementations of components to
their identifiers.  For the developer this is convenient, since they
only need to provide implementation terms while the framework takes
care of inter-component communication as specified in the
declarations. From the point of view of the framework, it provides
more control over the implementation: before execution static checks
can be done to determine if the terms provided by the application
developer conform to the specifications.

Declaring a component $C$ adds a case to the \texttt{implement}
macro. Now, a developer can use the form \texttt{(implement $C$ $f$)}
to bind a lambda function $f$ as the implementation of $C$. However,
not just any $f$ may be provided, as the arguments to
\texttt{implement} are subject to a Racket function
contract~\cite{dimoulas2011correct}.  Unfortunately there is a name
conflict between interaction contracts for components (as in DiaSuite) and function
contracts in Racket, which are not the same thing.  Function contracts
in Racket are flexible annotations on definitions and module exports, which
perform arbitrary tests at run-time on the input and output of functions.  For
example, a function can be annotated with a contract ensuring it maps
integers to integers. If the function receives or produces a
non-integer, the contract will trigger an error.  The contract on $f$
is derived from the interaction contracts of
Fig.~\ref{fig:racket-spec} as follows.

\paragraph*{Activation conditions}
These define the first argument to the function $f$.

\noindent\textbf{\texttt{when-provided \emph{x}}}.  First argument
gets type of $x$. For \texttt{\emph{Com\-pose\allowbreak Dis\-play}},
the contract starts with \texttt{(-> bitmap\%? \ldots)},\footnote{In
reality, \texttt{bitmap\%?} is shorthand for \texttt{(is-a?/c
bitmap\%)}, the contract builder which checks that a value is an
object of type \texttt{bitmap\%}.} since it is activated by
\texttt{\emph{Process\allowbreak Picture}} publishing a bitmap image.

\noindent\textbf{\texttt{when-required}}.
  No argument added -- the context was activated by pull.

\paragraph*{Data sources and actions}
These determine the (optional) next argument to the developer's function. This is a
closure providing proxied (that is, surrounded by a run-time guard) access to
the resource.  This makes it convenient for a developer to query a
resource, and allows the framework to enforce permissions.  Actions
for controllers are provided using the same mechanism.

\noindent\textbf{\texttt{get \emph{x}}}. The contract of the closure becomes
  \texttt{(-> t?)} where $t$ is the output type of $x$. 
  Note that there is no parameter, just a return value. This means that a component
requesting a value from another cannot exfiltrate data this way.
The full contract so far is therefore \texttt{(-> \ldots\ (-> t?) \ldots)}.

\noindent\textbf{\texttt{do \emph{x}}}. The contract of the closure becomes
  \texttt{(-> t? void?)} where $t$ is the input type of $x$. The full
  contract is therefore  \texttt{(-> \ldots\ (-> t? void?) void?)}. The final \texttt{void?}
reflects that controllers do not return values. 

\paragraph*{Publication requirements} These determine the last
arguments to the function contract of a context, corresponding to the output
type of the context. Publishing is handled using continuations, to give us
flexibility in the number of ``return'' statements provided.

\noindent\textbf{\texttt{always\_publish}}. One continuation function
corresponding to publication:     the
final contract becomes \texttt{(-> \ldots\ (-> t? void?) none/c)}, with $t$ the expected return type.

\noindent\textbf{\texttt{maybe\_publish}}. Two continuations to $f$,
for publish and no-publish. The first has the contract
\texttt{(-> t? void?)} with $t$ the output type. The second continuation simply returns control
to the framework.  If the developer
chooses not to publish, they use the second, no-publish continuation.
The contract is therefore \texttt{(-> \ldots\ (-> t? void?) (-> void?) none/c)}.

The \texttt{none/c} contract 
accepts no values: this causes a run-time error if the developer does not use one of the provided continuations.

\subsection{The  implementation of the application}
\label{sec:illustr-trans}

\begin{figure}
  \centering
\begin{lstlisting}[language=racket]
;;; Implementation file, webcamimpl.rkt
#lang s-exp "webcamspec.rkt"
(implement ComposeDisplay
  (lambda (pic getAdTxt publish nopublish)
    (let* ([canvas  (make-bitmap pic ..)]
           [adTxt   (getAdTxt)])
      (cond [(string=? "" adTxt) (nopublish)]) !\label{code:nopub}!
      ; ... do magic, overlay adTxt on pic
      (publish canvas)))) !\label{code:publish-composite}!
!\ldots! ; the remaining implement-terms
\end{lstlisting}  
  \caption{The implementation of the \texttt{ComposeDisplay} context.}
  \label{fig:racket-impl}
\end{figure}

In Fig.~\ref{fig:racket-impl}, we show a developer's possible
implementation of the context \texttt{ComposeDisplay}, which composes
the modified image with the advertisement text. Essentially, a
developer uses \texttt{\textbf{implement}} to bind their
implementation to the identifier introduced in the specifications, \cf
Fig.~\ref{fig:racket-spec}. Their implementation should be a lambda
term which obeys the contract resulting from the specification.  For
example, the \texttt{Compose\-Display} context has the contract
\texttt{(-> bitmap\%? (-> string?) (-> bitmap\%? void?) (-> void?)
  none/c)}%
. This is because it is activated by \texttt{Pro\-cess\-Pic\-ture}
publishing an image, it has \texttt{\textbf{get}}-access to the
\texttt{Make\-Ad} component which returns a string, and it may optionally
publish an image on account of its
\texttt{\textbf{may\-be\_\allowbreak pub\-lish}} specification.  The
last two arguments correspond to publishing \texttt{(-> bitmap\%?
  void?)} and not publishing \texttt{(-> void?)} continuations.  The lambda function
provided by the developer in Fig.~\ref{fig:racket-impl} conforms to
this contract. We see that if the advertisement component returns an
empty string (line~\ref{code:nopub}) the developer decides not to
publish, but otherwise the string is overlaid on the picture and the
developer publishes the composite image
(line~\ref{code:publish-composite}).

\begin{figure}
  \centering
  \includegraphics[angle=-90,width=0.8\linewidth,trim=0 0 4cm 0,clip=true]{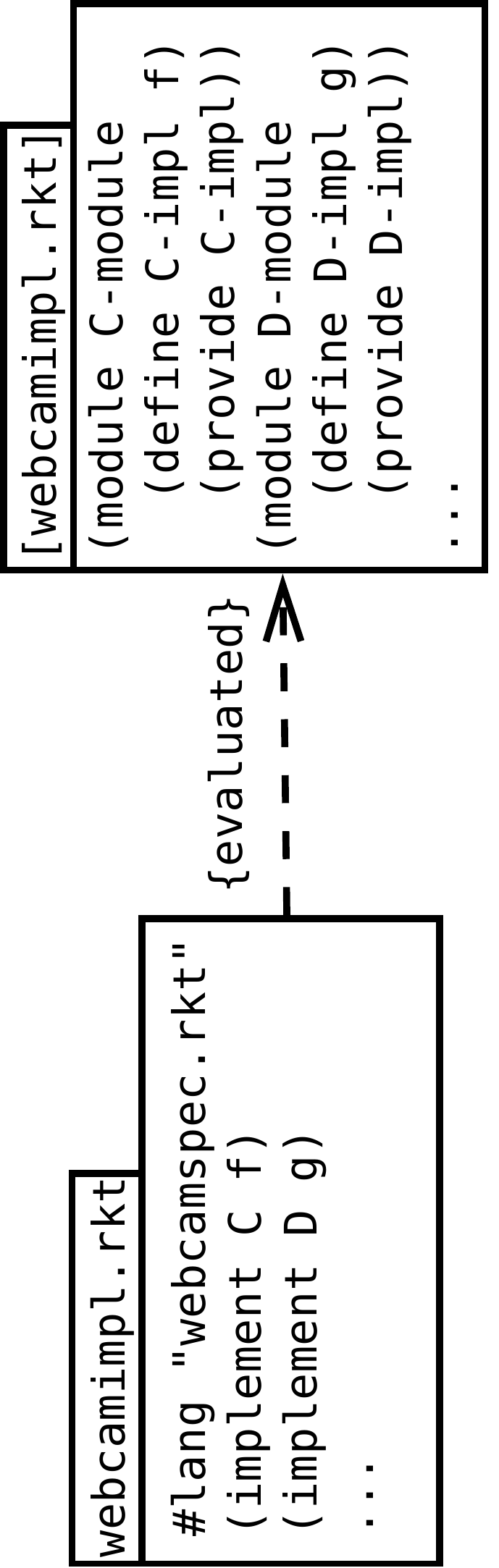}
  \caption{Separation of components using modules. The developer's
code (left), and its expanded form (right). $f$ in $C$ cannot access $D$ or $g$,
because of lexical scoping.}
  \label{fig:racket-submodules}
\end{figure}

To prevent implementations of different components communicating
outside of the condoned pathways, the \texttt{implement} macro wraps
each $f$ in its own submodule. As illustrated in
Fig.~\ref{fig:racket-submodules}, due to lexical scoping these do not
have access to surrounding terms, but merely export the implementation
for use in the top-level module.  The result of this wrapping is shown
in Fig.~\ref{fig:racket-impl-extended}. The code in grey is precisely
the term provided in Fig.~\ref{fig:racket-impl}, but it is now
isolated from the implementations of the other components, preventing
the developer from accessing them, which would constitute a leak.

\begin{figure}
  \centering
\begin{lstlisting}[language=racket,commentstyle=]
(module webcamimpl "webcamspec.rkt"
  (module ComposeDisplay-module racket/gui !\label{code:lang-spec}!
    (define/contract ComposeDisplay-impl
      (-> bitmap%? (-> string?) (-> bitmap%? void?)  !\label{code:contract}!
          (-> void?) none/c)<<@<<
      (lambda (pic getAdTxt publish nopublish)
        (let* ([canvas  (make-bitmap pic ..)]
               [adTxt   (getAdTxt)])  !\label{code:useproxy}!
          (cond [(string=? "" adTxt) (nopublish)])
          ; .. do magic, overlay adTxt on pic
          (publish canvas)))>>@>>)
    (provide ComposeDisplay-impl))
  !\ldots!)
\end{lstlisting}  
  \caption{The developer's code snippet is transformed into a submodule,
as a result of evaluating Fig.~\ref{fig:racket-impl}. The shaded code is simply 
the term the developer provided in Fig.~\ref{fig:racket-impl}.}
  \label{fig:racket-impl-extended}
\end{figure}

Note that alongside this snippet, the rest of the implementations of
the declared components must be provided in one module. This module
must be implemented using the new \texttt{webcamspec.rkt} language --
the one arising from the specifications we have written.  The
implementation module is checked before run-time to contain exactly
one \texttt{(implement $C$ \ldots)} term for each declared $C$.
However, we focus on this single context implementation to illustrate
what transformations are done on the developer's code.

When the developer has provided implementations for each of the
declared components, they can use the \texttt{(run)} convenience
function which is also exported by the module resulting from the
specifications. 
In the next section, we illustrate how these macros function.

\section{The framework and run-time}
\label{sec:framework}

Now that we have seen the user interface (\ie that which the application developer deals with) for our framework,
we elucidate how the framework is implemented. This is broken down into a number
of main parts:
(1)~the operation of the \texttt{define-context} and \texttt{define-controller} macros,
(2)~the expansion of the \texttt{implement} macro, and 
(3)~how the run-time support libraries tie the
implementations together to provide a coherent system.
These mechanisms are explained globally 
here, though certain implementation details are elided. 
Notably, getting all the needed identifiers we had introduced to be available in the right
transformer phases and module scopes was complicated. We invite the reader to experiment with the prototype code
-- the functionality for (1) and (2) is in
the \texttt{diaspec.rkt} module, the run-time library can be found in the
\texttt{fwexec.rkt} module.

\subsection{What happens with the declarations?}
\label{sec:work-define-cont}

\begin{figure}
  \centering
\begin{lstlisting}[language=racket]
(module webcamspec "diaspec.rkt" !\label{code:module-line}!
 (define ComposeDisplay !\label{code:compdisp-binding}!
   (context 'ComposeDisplay
     (interactioncontract ProcessPicture MakeAd
       'maybePublish) 'pic))
 (provide ComposeDisplay)
 (module+ contracts !\label{code:contract-submodule}!
   (define ComposeDisplay-contract
       (-> bitmap%? (-> string?)  !\label{code:contract-in-mod}!
           (-> bitmap%? void?) (-> void?) none/c))
   (provide ComposeDisplay-contract))
 (define-struct/contract ComposeDisplay-struct !\label{code:special-struct}!
   ([spec   (or/c context? controller?)]
    [implem (-> !$\ldots$!)])) ; contract from line!~\textit{\color{OliveGreen}{\ref{code:contract-in-mod}}}!
 (provide ComposeDisplay-struct
          implement-ComposeDisplay)
 (define-syntax (implement-ComposeDisplay stx) !\label{code:implement-macro}!
  (syntax-case stx (implement-ComposeDisplay)
   [(_ f) !\label{code:bind-f}!
     #'(begin
       (module ComposeDisplay-submodule racket/gui
         (require (submod "webcamspec.rkt" contracts)) !\label{code:import-contracts}!
         (provide ComposeDisplay-impl)
         (define/contract ComposeDisplay-impl
                          ComposeDisplay-contract f))
       (require (submod "." ComposeDisplay-submodule)) !\label{code:req-submodule}!
       (set-impl 'ComposeDisplay ; add to hashmap !\label{code:add-hashmap}!
         (ComposeDisplay-struct ComposeDisplay
            ComposeDisplay-impl)))]))
   
 (provide run (rename-out
                (module-begin-inner #%module-begin)))
 (define-syntax (module-begin-inner stx2)
    !$\ldots$! )) ;; omitted
\end{lstlisting}
  \caption{The simplified expansion of the specifications, concentrating on \texttt{ComposeDisplay} from Fig~\ref{fig:racket-spec}.
This code corresponds to the blue box in Fig.~\ref{fig:racket-global}.}
  \label{fig:expansion-of-specs}
\end{figure}

Previously we saw that the first step for a developer is to
declare the components of their application using the \texttt{de\allowbreak fine-\allowbreak con\allowbreak text} and
\texttt{de\allowbreak fi\allowbreak ne-\allowbreak con\allowbreak tro\allowbreak ller} keywords.  The specification should be provided
in a file which starts with a \texttt{\textbf{\#lang} s-\allowbreak exp "dia\-spec\-.rkt"} stanza, 
which causes the entire syntax tree of the specification to be passed to the
function exported from \texttt{diaspec.rkt} as \inlineracket$#
function does pattern matching on the specifications, and passes all occurrences of 
\texttt{define-} keywords to two handlers: (1)~to compute and store the associated contracts,
and (2)~to instantiate a struct which will later store the implementation. The introduced
identifiers are also stored as a list in the syntax transformer environment, \cf the ``Persistent effects''
system presented in~\cite{Tobin-hochstadt07advancedmacrology}. This compile-time storage
will later be used to check implementation modules: have all components been implemented,
and are all the identifiers used in the implementation declared in the specification?

To illustrate, Fig.~\ref{fig:expansion-of-specs} shows the expansion
of the \texttt{Com\-pose\-Dis\-play} declaration, from
line~\ref{code:compose-display-spec} of
Fig.~\ref{fig:racket-spec}. Simplifications have been
made, and module imports \etc have been omitted for brevity.
Some elements which are not specific to this declaration term have been elided,
namely a helper macro which transforms \texttt{(implement x ..)} terms into
\texttt{(implement-x ..)}, to correspond with the generated macro in line~\ref{code:implement-macro},
and a function which checks that all declared components have a 
corresponding \texttt{implement} term.
Finally, we also omit the generated syntax for \inlineracket$module-begin-inner$
from the specifications, since it is not particularly instructive. Note that it is this
definition which allows the implementation module to use the specification module as
its language, with the 
\inlineracket$#lang$ 
\texttt{s-\allowbreak exp "web\allowbreak cam\allowbreak spec\allowbreak .rkt"} directive.

Line~\ref{code:module-line} marks the start of the implementation module, called
\texttt{webcamspec}. It still references \inlineracket$"diaspec.rkt"$, which is 
the language the specification was written in, \cf Fig.~\ref{fig:racket-spec}.
This leaves us with the code resulting from the \texttt{Com\-pose\-Dis\-play}
context. In line~\ref{code:compdisp-binding}, we see that a binding is
introduced, using the name the developer chose for the component. Its 
value is a representation of the declaration, and is used to derive the contract.
In line~\ref{code:contract-submodule}, a submodule is appended with the Racket
contract the implementation is expected to adhere to. The \inlineracket$module+$
keyword adds terms to a named submodule, creating the submodule if necessary~\cite{flatt2013submodules}.
Line~\ref{code:special-struct} defines a tailored struct: it will hold the implementation
of \texttt{ComposeDisplay}, in the field tagged with the corresponding contract.
It becomes more interesting in line~\ref{code:implement-macro}, where we see that 
the \texttt{implement} keyword wraps the developer's implementation
in an independent submodule, as explained previously. This submodule will not have access to the
surrounding scope, hence the need for the \texttt{contracts} submodule, which 
we import in line~\ref{code:import-contracts}. 
As an aside, the \inlineracket$#'$ form is shorthand for \inlineracket$syntax$, which is
similar to \inlineracket$quote$, but produces a syntax object decorated
with lexical information and source-location information
that was attached to its argument at expansion time.
Crucially, it also substitutes \texttt{f}, the pattern variable bound by 
\inlineracket$syntax-case$ in line~\ref{code:bind-f}, with the pattern variable's match result, 
in this case the developer's implementation term.
Next, in line~\ref{code:req-submodule}, we have left the scope of the submodule.
We \inlineracket$require$ the submodule, bringing \texttt{ComposeDisplay-impl} into scope, which
we add to a hash map (line~\ref{code:add-hashmap}). This hash map associates names of components to their implementations.
Note how we are using the previously-defined struct, which forces the 
implementation term to adhere to its contract.

As a side-effect, \texttt{run} is only available to the developer if
they manage to evaluate the implementation module without compile
errors, which implies that only valid specifications and
implementations allow the developer to execute the framework. 
Since the implementation of the framework 
run-time library is mundane, we do not discuss it here. 
To run this code, \texttt{racket-mode}\footnote{Tested using MELPA version 20150330.1125
of Greg Hendershott's wonderful Racket mode for Emacs.} or DrRacket\footnote{Tested using DrRacket
v6.1.1.}~\cite{DBLP:journals/jfp/FindlerCFFKSF02} can be used. Simply
load the \texttt{webcamimpl.rkt} file, and when it is loaded, evaluate
\texttt{(run)} in the REPL.

\section{Evaluation and discussion}
\label{sec:discussion}

In the end, the application developer is presented with a reasonably
polished and coherent system for implementing an application in two
stages, which allows the platform to give the user more insight into
what mischief an application could potentially get up to. This assumes
that the specifications are distributed with the application, and
presented to the user (optionally formatted like
Fig.~\ref{fig:schematic-application}), and that the software is
compiled locally, or on a server that the user trusts. This would
ensure that the implementation does indeed conform to the
specifications.

We observe that going beyond Racket the functional
programming language, and using it as a language-building platform, is
where it really shines. We can mix, match and create languages as best
fits the niche, then glue modules together via the common
run-time library provided by Racket. This allows great flexibility and
control, since with Racket's \inlineracket$#lang$ mechanism, we can
precisely dictate the syntax and semantics of our new languages. These two
aspects therefore give Racket a lot of potential in the emerging
domain of declaration-driven frameworks.

\subsection{Limitations}
\label{sec:limitations}

Unfortunately, there are issues that would need to be resolved before the
proposed approach would be feasible in the real world. One of the trickiest 
parts of ensuring no communication between components is that consequently 
we cannot allow a developer to use any external modules in their code. This
is because if a developer could \texttt{\textbf{require}} any module,
they could in effect execute arbitrary code. It could also be used as a
communication channel, since modules have mutable state. 
Therefore,
in the prototype, we chose not to allow any importing of modules, but for a realistic
application this would probably not be acceptable  -- we could imagine
needing to use a library for parsing JSON, or processing images, or any number of 
benign tasks. Perhaps this would be a decision for the platform provider to make: 
is a particular library ``safe'' and could it be white-listed?

Another potential leak could be the \texttt{\textbf{eval}} form.
Using it, a developer could easily obfuscate any behaviour desired. In
fact, arbitrary imports and calls would be possible that way.
We therefore inspect a developer's implementation for such things as
the use of \texttt{\textbf{eval}}, and reject them syntactically, but
since the binding might be hidden or renamed, this approach is not
necessarily robust.  This highlights a need in Racket: allowing
components or functions to be pure would solve this
vulnerability. Perhaps Typed
Racket~\cite{Tobin-hochstadt07advancedmacrology} will offer a solution
in the future -- purity analysis is on the project to-do
list.\footnote{The page ``Typed Racket Plans'' at
  \url{https://github.com/plt/racket/wiki/Typed-Racket-plans} gives us
  hope.}

It was also rather finicky to implement all the macros as described above. Although 
conceptually simple, it turns out to be pretty difficult in practice to get all the
identifiers to be available in the right syntax transformer phases.
The macro debugger in DrRacket is
quite powerful, but unfortunately still leaves a lot to be desired. For example,
we failed to get it to show the completely expanded implementation module as it is 
presented in Fig.~\ref{fig:expansion-of-specs} -- that code is largely
worked out by hand. Finer control over the macro unfolding would
therefore be beneficial.

The end result is not at all pessimistic, in spite of these
shortcomings and difficulties. The prototype does demonstrate the
power of a language such as Racket, which gives a programmer the
capability to easily modify syntax and provide custom
interpretations. This prototype also demonstrates that it is possible
to cleanly separate concerns and enforce a certain structure on the
final implementation.

\subsection{Future work} 
\label{sec:future-work-}

There are a number of clear avenues for improving this work. Firstly,
we note that the chosen platform and model are merely examples, it
should be easy to build similar ``active'' specification DSLs for
other domains.  This modular approach is also very flexible: we could
choose to use any Racket extension as the implementation language for
the developer to use, whether it be FRTime or Typed Racket or any
other of the many libraries. We could even decide to provide different
languages for different modules -- the changes would be minor. If for
example Typed Racket were to support purity analysis in the future,
this would be a very attractive option, allowing us to be confident that
no unwanted communication between components is possible. As stated, though,
before this approach could be introduced into the wild, a safe module
importing mechanism should be devised.

Another aspect to be dealt with is a very practical one: how to
integrate this approach into an application store, where users could
download applications for use on their local platforms. As it stands,
the developer would have to submit their specification and
implementation modules as source code, and the application store would
need to compile them together, to ensure that the contracts and
modules have not been tampered with. The application store -- which
the user would have to trust -- could then distribute compiled
versions of the application which would be compatible with the
run-time library locally available on users' devices. Clearly, this is
not be desirable in all situations: most commercial application
developers submit compiled versions of their software, which in our
case could allow them to \eg modify the contracts, rendering the
applications unsafe.

\section{Conclusion}
\label{sec:conclusion}

In conclusion, we have tried to address the problem of resource access
control to protect privacy of end users' sensitive data. Taking
inspiration from the DiaSuite approach, we demonstrate an embedded DSL
for specifying applications, which itself unfolds to a programming
environment, placing restrictions on the application developer. While
the prototype is not a perfect solution to the problem, it does
demonstrate a novel approach to resource control which is very
versatile, by nature of being entirely composed of embedded DSLs. It
also offers users more insight into what sensitive resources are used
for, compared to currently widespread mobile platforms.

In the future, it would be interesting to explore the use of other Racket
libraries, particularly Typed Racket, in the hope that we can achieve more 
reliable restrictions than currently possible. This might be an avenue
to pursue in response to the vulnerability that the current prototype
has, arising from evaluation of dynamically constructed expressions or
allowing module importing.

\section*{Acknowledgements} 

Special thanks go to 
Ludovic Court\`es,   
Camille Ma\~nano,    
Andreas Enge,        
and 
Hamish Ivey-Law,     
who proofread early drafts of this work and provided invaluable
comments. The constructive criticism
provided by the anonymous reviewers is equally appreciated.



\end{document}